\documentstyle[preprint,aps,epsfig]{revtex}

\newcommand{\lsim}{\raisebox{0.3mm}{\em $\, <$} 
\hspace{-3.3mm} \raisebox{-1.8mm}{\em $\sim \,$}}
\newcommand{\gsim}{\raisebox{0.3mm}{\em $\, >$}
\hspace{-3.3mm} \raisebox{-1.8mm}{\em $\sim \,$}}

\begin{document}
\baselineskip 6.7mm
\begin{flushright}
\vglue -1.0cm
hep-ph/0111269 \\ 
\end{flushright}

\draft
\begin{center}
\Large\bf
Lower Bound on $|U_{e3}|^2$ from Single and Double Beta Decay 
Experiments
\end{center}
\vskip 0.5cm
\begin{center}
Hisakazu Minakata$^{1,2}$ 
\footnote[1]{E-mail: minakata@phys.metro-u.ac.jp}
and Hiroaki Sugiyama$^{1}$
\footnote[2]{E-mail: hiroaki@phys.metro-u.ac.jp}\\
\vskip 0.2cm
{$^1$\it Department of Physics, Tokyo Metropolitan University \\
1-1 Minami-Osawa, Hachioji, Tokyo 192-0397, Japan}
\vskip 0.2cm
{$^2$\it Center for Theoretical Physics, 
Massachusetts Institute of Technology \\
Cambridge, MA 02139, USA}

\end{center}
\date{October 2001}

\vspace{-1.0cm}

\begin{abstract}
We point out under the assumption of Majorana neutrinos 
that a lower bound on the MNS matrix element 
$|U_{e3}|^2$ can be derived by using constraint imposed by 
neutrinoless double beta decay experiments and by positive 
detection of neutrino mass by single beta decay experiments. 
We show that the lower bound exists in a narrow region of 
the ratio of the observables in these two experiments, 
$\langle m \rangle_{\beta \beta}/\langle m \rangle_{\beta}$.  
It means that once the neutrino mass is detected in the 
bound-sensitive region one must soon observe signal 
in neutrinoless double beta decay experiments.

\end{abstract}

\vskip 1cm

\pacs{14.60.Pq, 26.65.+t, 23.40.-s}

\newpage
\section{Introduction}

There exist, by now, accumulated evidences in the atmospheric 
\cite {SKatm} and the solar neutrino \cite {solar} observations 
that neutrinos do oscillate. 
The existence of neutrino oscillation is further strengthened by 
the result of the first long-baseline man-made beam experiment K2K, 
in particular by their latest result \cite{K2K}.
They constitute the first compelling evidence for physics 
beyond the standard model of particle physics. In particular, we have 
leaned that an almost maximal mixing angle $\theta_{23}$ is required 
to account for the atmospheric neutrino anomaly, which is quite 
unexpected from our experience in the quark sector. 

What is even more surprizing to us is that, according to the latest 
global analyses of the current solar neutrino data 
\cite {bari,goswami,jnb-concha1,KS01,jnb-concha2}, 
the angle $\theta_{12}$ which is responsible for the solar neutrino 
oscillation is likely to be large though may not necessarily be maximal. 
It is in sharp contrast with the fact that the remaining mixing angle 
$\theta_{13}$ is constrained to be small, $s^2_{13} \lsim 0.03$, 
by the reactor experiments \cite {CHOOZ}. 
Given the current status of our understanding of the structure of 
lepton flavor mixing matrix, the MNS matrix \cite{MNS}, it would 
be nice if there are any hints on how small is the angle 
$\theta_{13}$.\footnote{
See e.g., \cite{NOW2000mina} for a summary of remaining issues in 
three flavor mixing scheme of neutrinos.} 

In this paper, we try to pursue such a possibility and point out 
that one can derive a lower bound on $s^2_{13} = |U_{e3}|^2$ 
through joint efforts by double beta decay experiments and 
by direct mass determination either by single beta decay or 
by cosmological observations. 
We assume in this paper that neutrinos are Majorana 
particle to rely on the bound imposed by double beta decay 
experiments. 

Before getting into the bussiness, let us briefly summarize 
existing knowledge on how $\theta_{13}$ can be measured, or 
further constrained. Most optimistically, the next 
generation long baseline experiments, MINOS \cite{MINOS}, 
JHF \cite{JHF}, and OPERA \cite{OPERA} will observe $\nu_e$ 
appearance events and measure the angle $\theta_{13}$. 
Most notably, the JHF can probe 
$\sin^2{2 \theta_{13}} \gsim$ several $\times 10^{-3}$ 
in its phase I \cite{JHF}. 
If realized, a large-volume reactor experiment \cite{krasnoyarsk}
can also probe the similar (to a slightly shallower) region. 
If the angle is smaller than the sensitivity region of these 
experiments, we have to wait for future supermassive detector 
experiments, utilizing either low energy conventional superbeams 
or neutrino factoties. 
(See e.g., \cite {MNjhep01} for references cited therein.) 
If nature is so unkind as to tune the angle extremely small, 
$\sin^2{2 \theta_{13}} \ll 10^{-5}$, then the only way to detect its 
effect would be via supernova neutrinos \cite {SNnu}.

\section{Constraint from neutrinoless double beta decay}

Let us start by examining constraint from double beta decay.
We use throughout this paper the standard notation of the MNS matrix:
\begin{equation}
U=\left[
\begin{array}{ccc}
c_{12}c_{13} & s_{12}c_{13} &   s_{13}e^{-i\delta}\nonumber\\
-s_{12}c_{23}-c_{12}s_{23}s_{13}e^{i\delta} &
c_{12}c_{23}-s_{12}s_{23}s_{13}e^{i\delta} & s_{23}c_{13}\nonumber\\
s_{12}s_{23}-c_{12}c_{23}s_{13}e^{i\delta} &
-c_{12}s_{23}-s_{12}c_{23}s_{13}e^{i\delta} & c_{23}c_{13}\nonumber\\
\end{array}
\right].
\label{MNSmatrix}
\end{equation}
Using the notation, the observable in neutrinoless double beta decay 
experiments can be expressed as 
\begin{eqnarray}
\langle m \rangle_{\beta \beta}
&=& 
\left\vert 
\hskip 0.2cm
\sum^{3}_{i=1}
m_i U^2_{ei} 
\hskip 0.2cm
\right\vert \nonumber\\
&=&
\left\vert 
\hskip 0.2cm
m_1 c_{12}^2c_{13}^2 e^{-i \beta}
+ m_2 s_{12}^2c_{13}^2 e^{+i \beta}
+ m_3 s_{13}^2 e^{i(3 \gamma - 2\delta)}
\hskip 0.2cm
\right\vert,
\label{beta1}
\end{eqnarray}
where $m_i$ (i=1, 2, 3) denote neutrino mass eigenvalues, $U_{ei}$ 
are the elements in the first low of the MNS matrix, and 
$\beta$ and $\gamma$ are the extra CP-violating phases 
characteristic to Majorana neutrinos \cite {SV,FY}. 
We have used in the second line of (\ref{beta1}) the Majorana phases 
in the convention of \cite {MY97}.
There have been large number of papers quite recently which 
devoted to extract constraints from neutrinoless double beta 
decay experiments \cite {MY97,dblbeta1,dblbeta2}. 

We define the neutrino mass-squared difference as
$\Delta m^2_{ij} \equiv m^2_{j} - m^2_{i}$ in this paper.
In the following analysis, we must distinguish the two 
different neutrino mass patterns, the normal ($\Delta m^2_{23}>0$) 
vs. inverted ($\Delta m^2_{23}<0$) mass hierarchies. 
We use the convention that $m_3$ is the largest (smallest) mass 
in the normal (inverted) mass hierarchy so that the angles 
$\theta_{12}$ and $\theta_{23}$ are always responsible for the 
solar and the atmospheric neutrino oscillations, respectively. 
We therefore sometimes use the notations 
$\Delta m^2_{23} \equiv \Delta m^2_{atm}$ and 
$\Delta m^2_{12} \equiv \Delta m^2_{\odot}$ to emphasize 
that they are experimentally (the latter to be) measured quantities. 
Because of the hierarchy of mass scales, 
$\Delta m^2_{\odot}/\Delta m^2_{atm} \ll 1$, $\Delta m^2_{12}$
can be made always positive as far as $\theta_{12}$ is taken in
its full range [0, $\pi/2$]\cite {FLMP}.

In order to derive constraint on mixing parameters we need the 
classification.

\noindent
Case A:
\begin{equation}
\left\vert m_1 c_{12}^2c_{13}^2 e^{-i \beta}
+ m_2 s_{12}^2c_{13}^2 e^{+i \beta}
\right\vert
\geq m_3 s^2_{13}
\end{equation}

\noindent
Case B:
\begin{equation}
\left\vert m_1 c_{12}^2c_{13}^2 e^{-i \beta}
+ m_2 s_{12}^2c_{13}^2 e^{+i \beta}
\right\vert
\leq m_3 s^2_{13}
\end{equation}
Both types of mass hierarchies are allowed in the cases A and B. 
For a given experimental upper bound on 
$\langle m \rangle_{\beta \beta}$, 
one can derive a lower (upper) bound on $s^2_{13}$ 
in the case A (B). Therefore, we start with the case A.

\subsection{Case A}

In this case, the lower bound on $\langle m \rangle_{\beta \beta}$ 
can be obtained as in the following way;

\begin{eqnarray}
\langle m \rangle_{\beta \beta} 
&\geq& c_{13}^2
\left\vert (m_1 c_{12}^2 + m_2 s_{12}^2) \cos{\beta}
- i (m_1 c_{12}^2 - m_2 s_{12}^2) \sin{\beta}
\right\vert
- m_3 s_{13}^2 \nonumber\\
&=&
c_{13}^2 \sqrt{m_1^2 c_{12}^4 + m_2^2 s_{12}^4 
+ 2 m_1 m_2 c_{12}^2 s_{12}^2 \cos{2 \beta}}
- m_3 s_{13}^2. 
\label{beta3}
\end{eqnarray}
Noticing that the right-hand-side of (\ref{beta3}) has a minimum 
at $\cos{2 \beta}=-1$, we obtain the inequality
\begin{equation}
\langle m \rangle_{\beta \beta} \geq
c_{13}^2 
\left\vert m_1 c_{12}^2 - m_2 s_{12}^2 \right\vert
- m_3 s^2_{13}.
\end{equation}

If a neutrinoless double beta decay experiment imposes 
the bound 
$\langle m \rangle_{\beta \beta} \leq \langle m \rangle_{\beta \beta}^{exp}$
the lower bound on $s^2_{13}$ results;
\begin{equation}
s^2_{13} \geq 
\frac {\left\vert m_1 c_{12}^2 - m_2 s_{12}^2 \right\vert - 
\langle m \rangle_{\beta \beta}^{exp}}
{\left\vert m_1 c_{12}^2 - m_2 s_{12}^2 \right\vert + m_3}.
\label{bound1}
\end{equation}

We rewrite the lower bound into the one expressed by the heaviest 
neutrino mass $m_H$ and $\Delta m^2$ measured by the atmospheric 
and the solar neutrino experiments. We note that 
the neutrino masses $m_i$ (i=1,2,3) can be parametrized by 
$\Delta m^2_{atm} = \Delta m^2_{23}$, 
$\Delta m^2_{\odot} = \Delta m^2_{12}$, 
and a absolute mass scale $m_H$. We take $m_H = m_3$ and 
$m_H = m_2$ for the normal and the inverted mass hierarchies, 
respectively. Then,
\begin{equation}
m_1 = \sqrt{m_H^2 - \Delta m^2_{atm} - \Delta m^2_{\odot}},
\hskip 1cm
m_2 = \sqrt{m_H^2 - \Delta m^2_{atm}},
\hskip 1cm
m_3 = m_H, 
\end{equation}
for the normal mass hierarchy, and 
\begin{equation}
m_1 = \sqrt{m_H^2 - \Delta m^2_{\odot}},
\hskip 1cm
m_2 = m_H, 
\hskip 1cm
m_3 = \sqrt{m_H^2 - |\Delta m^2_{atm}|},
\end{equation}
for the inverted mass hierarchy.

We express the lower bound on $s_{13}^2$ by using the dimensionless 
ratios
\begin{equation}
R_{\beta \beta} \equiv \frac{\langle m \rangle_{\beta \beta}^{exp}} {m_H},
\hskip 1cm
r_{atm} \equiv \frac{\sqrt{|\Delta m^2_{atm}|}} {m_H},
\hskip 1cm
r_{\odot} \equiv \frac{\sqrt{\Delta m^2_{\odot}}} {m_H}.
\end{equation}
The lower bound on $s_{13}^2$ reads for each type of mass 
hierarchy as follows:

\noindent
(i) Normal mass hierarchy; $m_H = m_3$

\begin{equation}
s_{13}^2 \geq
\frac{
\left\vert \sqrt{1 - r_{atm}^2 - r_{\odot}^2} c_{12}^2 -
\sqrt{1 - r_{atm}^2} s_{12}^2 \right\vert - 
R_{\beta \beta}
}
{\left\vert \sqrt{1 - r_{atm}^2 - r_{\odot}^2} c_{12}^2 -
\sqrt{1 - r_{atm}^2} s_{12}^2 \right\vert + 1}.
\end{equation}

\noindent
(ii) Inverted mass hierarchy; $m_H = m_2$

\begin{equation}
s_{13}^2 \geq 
\frac{
\left\vert \sqrt{1 - r_{\odot}^2} c_{12}^2 -
s_{12}^2 \right\vert - 
R_{\beta \beta}
}
{\left\vert \sqrt{1 - r_{\odot}^2} c_{12}^2 -
s_{12}^2 \right\vert + 
\sqrt{1 - r_{atm}^2}}.
\end{equation}

The parameter $r_{\odot}^2$ is extremely small, 
$r_{\odot}^2 = \Delta m^2_{\odot}/m_H^2 \lsim 10^{-3}$ for 
the possible best sensitivity of $\sim 0.3$ eV. (See later.)
Hence, it is an excellent approximation to ignore $r_{\odot}^2$ 
unless the sensitivity goes down to very close to $\Delta m^2_{atm}$
so that it is comparable with $1 - r_{atm}^2$ 
in the normal mass hierarchy case. 
Ignoring $r_{\odot}^2$ the lower bound on $s_{13}^2$ 
greatly simplifies:
\begin{equation}
s_{13}^2 \geq 
\frac{
|\cos{2 \theta_{12}}|
\sqrt{1 - r_{atm}^2} - 
R_{\beta \beta}}
{|\cos{2 \theta_{12}}|
\sqrt{1 - r_{atm}^2} + 1}
\hskip 1cm
\mbox {(normal mass hierarchy)},
\label{bound_normal}
\end{equation}
\begin{equation}
s_{13}^2 \geq
\frac{
|\cos{2 \theta_{12}}| - R_{\beta \beta}}
{|\cos{2 \theta_{12}}| + 
\sqrt{1 - r_{atm}^2}}
\hskip 1cm
\mbox {(inverted mass hierarchy)}.
\label{bound_inverted}
\end{equation}
We finally note that in the degenerate mass limit 
$r_{atm} \rightarrow 0$ the lower bound ceases to 
distinguish between the mass patterns, and has a universal form
\begin{equation}
s_{13}^2 \geq 
\frac{
|\cos{2 \theta_{12}}| - R_{\beta \beta}}
{|\cos{2 \theta_{12}}| + 1}
\hskip 1cm
\mbox {(degenerate mass limit)}.
\label{bound_degenerate}
\end{equation}

\subsection{Case B}

For completeness, we treat the case B, which yields the upper bound on 
$s_{13}^2$. Proceeding via the similar way toward (\ref{beta3}) 
we obtain the inequality 
\begin{equation}
\langle m \rangle_{\beta \beta} \geq 
m_3 s_{13}^2 - 
c_{13}^2 
\left\vert m_1 c_{12}^2 + m_2 s_{12}^2 \right\vert
\end{equation}
which is saturated at $\cos{2 \beta}=+1$. 
Then, the upper bound on $s_{13}^2$ for a given experimental 
bound on $\langle m \rangle_{\beta \beta}$ entails after 
ignoring $r_{\odot}^2$ as 
\begin{equation}
s_{13}^2 \leq 
\frac{\sqrt{1 - r_{atm}^2} + R_{\beta \beta}}
{\sqrt{1 - r_{atm}^2} + 1}
\hskip 1cm
\mbox {(normal mass hierarchy)},
\label{upperbound1}
\end{equation}
\begin{equation}
s_{13}^2 \leq 
\frac{1 + R_{\beta \beta}}
{1 + \sqrt{1 - r_{atm}^2}}
\hskip 1cm
\mbox {(inverted mass hierarchy)}.
\label{upperbound2}
\end{equation}
It can give a stronger bound than the CHOOZ limit only 
for the normal hierarchy and if $R_{\beta \beta} < 0.03$. 
We do not discuss it further because it is outside of 
our analysis in section IV.

\section{Connection between $m_H$ and the observable in 
direct mass measurements}

Before entering into the actual analysis of the bound, we clarify 
the relationship between the largest mass $m_H$ 
($= m_3$ for normal, and $= m_2$ for inverted mass hierarchies) 
and the observable in direct mass measurement in single beta decay 
experiments.\footnote{
For the present status and the future prospects of this type 
of experiments, see e.g., the website of a recent conference 
devoted to the topics \cite {sub-eV}.}
We argue that $m_H$ can be identified as the observable in such 
experiments in a good approximation. Suppose that the neutrino 
masses are hierarchical and obey either  
$m_3 \gg m_2 \simeq m_1$ (normal mass hierarchy), or 
$m_2 \simeq m_1 \gg m_3$ (inverted mass hierarchy). 
In the former case, it is obvious that $m_3 = m_H$ is the observable. 
In the latter case, it was shown in \cite {FPS01} that the observable 
in direct mass measurements $\langle m \rangle_{\beta}$ 
is given by 
\begin{equation}
\langle m \rangle_{\beta} = 
\frac{
\sum^{n}_{j=1}
m_j |U_{ej}|^2}
{\sum^{n}_{j=1} |U_{ej}|^2}
\label{FPS}
\end{equation}
where $n$ is the dimension of the subspace of (approximately) 
degenerate mass neutrinos, and $n=2$ in the case under discussion.
Then, $\langle m \rangle_{\beta} = m_2 = m_H$ in a good approximation.  
In the opposite extreme, $m_i^2 \gg \Delta m^2_{atm}$ which is usually 
referred to as "almost degenerate neutrinos", (\ref{FPS}) with $n=3$ 
tells us that $\langle m \rangle_{\beta} = m_H$ in a very good 
approximation.  

Thus, $\langle m \rangle_{\beta} = m_H$ holds in both extreme, 
hierarchical and degenerate mass neutrinos. 
This discussion strongly suggests that 
$\langle m \rangle_{\beta}$ is reasonably well approximated 
by $m_H$ even in the intermediate region.

\section{Analysis of the lower bound on $s_{13}^2$}

We present in Fig.1 the lower bound on $s_{13}^2$ as a 
function of $m_H$ for various values of $\cos{2 \theta_{12}}$; 
the region lower-right to each curve is excluded. The shaded area 
indicates the region excluded by the CHOOZ experiment \cite {CHOOZ}, 
which we approximate as $s_{13}^2 < 0.03$.\footnote{
While the precise value of the CHOOZ constraint actually depends 
upon the value of $|\Delta m^2_{atm}|$ \cite {CHOOZ}, we do not 
elaborate this point in this paper.}
We take two typical values of 
$\langle m \rangle_{\beta \beta}^{exp}$, 0.34 eV and 0.1 eV. 
We present in Fig. 1 only the results for the normal mass 
hierarchy. It is because the results barely changes in the 
case of inverted mass hierarchy; there is practically no 
diference in (a) 
$\langle m \rangle_{\beta \beta}^{exp} = 0.34$ eV case, and 
the curve shifts toward the right by about 4 \% in (b) 
$\langle m \rangle_{\beta \beta}^{exp} = 0.1$ eV case at the 
best fit value of LMA solution (see below).

We note that the former value of 
$\langle m \rangle_{\beta \beta}^{exp}$
corresponds to the present 90 \% CL bound by 
Heidelberg-Moscow group \cite{klapdor}, while the latter 
indicates a modest sensitivity to be achieved in near future 
by CUORE \cite {CUORE}, GENIUS \cite{GENIUS}, and by 
MOON \cite {MOON} double beta decay experiments.
We would like to remind the readers that the present strongest 
upper limit on  $\langle m \rangle_{\beta}$ is from the 
Mainz collaboration \cite {Mainz}, 
$\langle m \rangle_{\beta} \leq 2.2$ eV (95 \% CL).
A similar bound 
$\langle m \rangle_{\beta} \leq 2.5$ eV (95 \% CL) 
is derived by the Troitsk group \cite {Troitsk}.
The sensitivity of the proposed KATRIN experiment is 
expected to extend to 
$\langle m \rangle_{\beta} \leq 0.3$ eV \cite {KATRIN}. 

In the numerical analysis in this section we assume  
$|\Delta m^2_{atm}| = 3 \times 10^{-3}$ eV$^2$, the best fit value 
\cite{FLM01} of the combined data of Super-Kamiokande \cite {SKatm} 
and the K2K \cite {K2K} experiments. We ignore 
$\Delta m^2_{\odot}$ in most part of our analysis and its 
effect is detectable only in a limited region outside of 
the sensitivity of the KATRIN experiment, as shown in 
Fig. 2.

We note that the allowed region of the mixing angle $\theta_{12}$ 
has a large uncertainty. In particular, its value at the largest 
end is far more uncertain compared with the accuracy we need to 
determine where is the bound-sensitive region. 
We just quote for illustration the allowed region for the 
LMA solution given in \cite {KS01}; 
$\sim 0.67 \geq  
\cos {2 \theta_{12}} \geq 0.19$ (95 \% CL), $\geq 0.099$ (99 \% CL), and 
$\geq  - 0.024$ (99.73 \% CL). 
Then, the allowed region of the LMA solution extends even 
at 95 \% CL to the right-most contour in Fig. 1. 
Therefore, the Mainz and the Troitsk experiments begin to touch 
the parameter region already at their present sensitivities. 
Similarly, the 95 \% allowed regions of the LOW and the VAC 
solutions extend, very roughly to, 
$0.30 \geq  \cos {2 \theta_{12}}  \geq 0.026$ and  
$0.57 \geq  |\cos {2 \theta_{12}}| \geq  0.30$, 
respectively \cite {KS01}.

We observe in Fig. 1b that for 
$\langle m \rangle_{\beta \beta}^{exp} = 0.1$ eV 
the KATRIN experiment start to cover a large portion of 
the parameter region of the LMA solution toward its best fit 
value of $\cos {2 \theta_{12}} = 0.48 $ \cite {KS01}
(0.46 in \cite {jnb-concha2}).
Therefore, if the LMA solution is confirmed by the KamLAND 
reactor experiment and if the KATRIN experiment detects 
direct neutrino mass, we would hope that we will have a 
lower bound on $s_{13}^2$ at the same time.

In Fig. 2 we present the similar plot as in Fig. 1 but for 
$\langle m \rangle_{\beta \beta}^{exp} = 0.01$ eV. In this case 
the cases of the normal vs. the inverted mass hierarchies  
differ clearly from each other, as one can see by comparing 
Figs. 2a and 2b. 
%
In region of $m_{H} \sim \sqrt{|\Delta m^2_{atm}|} \simeq 0.06$ eV 
the lines of positive and negative $\cos{2 \theta_{12}}$ 
start to split by nonvanishing $r_{\odot}^2$ correction, 
as indicated by the dashed lines in Fig. 2.


The significant feature of the lower bound we have obtained 
is that it exists in a narrow window of $m_H$, the highest 
neutrino mass, whose values depend very sensitively on 
$\langle m \rangle_{\beta \beta}$, 
the observable in neutrinoless double beta decay experiment. 
We now try to characterize the region, the bound-sensitive region, 
as a function of the experimental upper limit 
$\langle m \rangle_{\beta \beta}^{exp}$ and 
$\langle m \rangle_{\beta}$.
In the degenerate mass limit 
$r_{atm} \rightarrow 0$, it can readily be done 
by using (\ref {bound_degenerate}). By demanding the consistency 
with the CHOOZ bound, we obtain 
\begin{equation}
0.97 |\cos{2 \theta_{12}}| - 0.03 \leq  
R_{\beta \beta} \simeq
\frac {\langle m \rangle_{\beta \beta}^{exp}}{\langle m \rangle_{\beta}} \leq 
|\cos{2 \theta_{12}}|.
\label{dmlimit}
\end{equation}
Therefore, the bound-sensitive region is characterized by 
the ratio of these two experimental observables in the 
degenerate mass approximation. This feature prevails to 
certain extent beyond the approximation as we will see.

In Fig. 3 we present plots of the bound-sensitive 
region expressed by the ratio 
$R_{\beta \beta} \equiv 
\langle m \rangle_{\beta \beta}^{exp}/\langle m \rangle_{\beta}$ 
as a function of $\cos{2 \theta_{12}}$ for the normal and the 
inverted mass hierarchies in regions for 
(a) LMA and (b) LOW solar neutrino solutions. 
The region covered is determined as the best fit 
value $\pm$ 10 \% in $\tan^2{\theta_{12}}$. For the LMA solutions 
it roughly corresponds to the accuracy to be achieved in 
KamLAND \cite {MP01}. We do not present the case of vacuum 
solution because the relevant region roughly 
overlaps with that of LMA solution.
In the degenerate mass limit we must have a universal curve 
(\ref{dmlimit}) and the splitting between the normal and the inverted 
hierarchy cases in Fig. 3 represents corrections by the effect of 
nonvanishing $r_{atm}$. 
We do not plot the scaling curve (\ref{dmlimit}) because it is 
virtually identical with the one for the inverted 
mass hierarchy.

Notice that the ratio $R_{\beta \beta}$ can be small for each mass 
pattern because of the suppression of contribution 
from highest mass neutrinos by small $s_{13}^2$ in the case 
of normal mass hierarchy, and by possible cancellation in the 
inverted mass hierarchy.
Once $R_{\beta \beta}$ is given, it is easy for the readers 
to read off the lower bound on $s_{13}^2$ for a given value 
of $\theta_{12}$ because the function of lower bound 
(\ref{bound_normal}) and 
(\ref{bound_inverted}) are approximately linear 
within the narrow strip $0 \leq s_{13}^2 \leq 0.03$.

We note, however, that several highly nontrivial 
requirements must be met in order to extract the lowest 
value of $s_{13}^2$.
We see from fig. 3 that the accuracy of direct mass 
measurement of $\langle m \rangle_{\beta}$ at given 
$\langle m \rangle_{\beta \beta}^{exp}$ 
must be better than roughly $\pm$ 5 \% for 
the LMA and $\pm$ 20 \% for the LOW solutions, respectively. 
Furthermore, the precise values of mixing parameters, in particular, 
$\cos{2 \theta_{12}}$ must be known to better than $\pm$ 10 \% level.

Finally, some remarks are in order:

\vskip 0.3cm

\noindent 
(1) Suppose that in a future time the solar mixing angle $\theta_{12}$ 
is determined by some ingenious experiments. 
Assume then that we have obtained the lower bound on $s_{13}^2$ 
by KATRIN observation of neutrino mass at the time of sensitivity 
$\langle m \rangle_{\beta \beta}^{exp} = 0.1$ eV of double 
beta decay experiment. 
Now, further suppose that the latter experiment would have been 
improved so that 
$\langle m \rangle_{\beta \beta}^{exp} \ll 0.1$ eV. Figure 3 
tells us that then we are in the excluded region. 
What does it mean?

The answer is; it means that the double beta decay experiment 
must see positive events before going down to 
$\langle m \rangle_{\beta \beta}^{exp} \ll 0.1$ eV. 
If not, it is the indication that nature chooses a different 
lepton flavor mixing scheme from the standard one, or 
our assumption of Majorana neutrinos is wrong. The last possibility 
is a rare case of excluding Majorana hypothesis in spite of finite 
resolution of $\langle m \rangle_{\beta \beta}$ measurement, which  
takes place owing to the CHOOZ constraint.
It is very exciting that a discovery in an experiment 
either signals discovery of another experiment, 
or indicate radically different views of lepton sector from 
the one useally accepted.

\vskip 0.3cm

\noindent 
(2) We have explored in this paper features of the lower bound on 
$s_{13}^2$
which can be derived by using constraints imposed by the single 
and the double beta decay experiments. The similar consideration 
can be done to constrain different parameters such as 
solar mixing angle $\theta_{12}$, for example. 
A more generic analysis of the bound is in progress \cite {MS02}.

\vskip 0.3cm

\noindent 
(3) While we have focused on beta decay experiments in this paper, 
the possibility of obtaining better knowledge of neutrino mass 
in terms of cosmological observation must be pursuit. See e.g., 
\cite {fukugita}.

\acknowledgments 

We thank Theoretical Physics Department of Fermilab for warm 
hospitality extended to us where this work has been initiated. 
HM thanks Center for Theoretical Physics of MIT for hospitality 
where this work was completed. 
This work was supported by the Grant-in-Aid for Scientific Research 
in Priority Areas No. 12047222, Japan Ministry 
of Education, Culture, Sports, Science, and Technology.


\begin{figure}[ht]
\vglue 1.0cm
\hglue -1.0cm 
\centerline{\protect\hbox{
\psfig{file=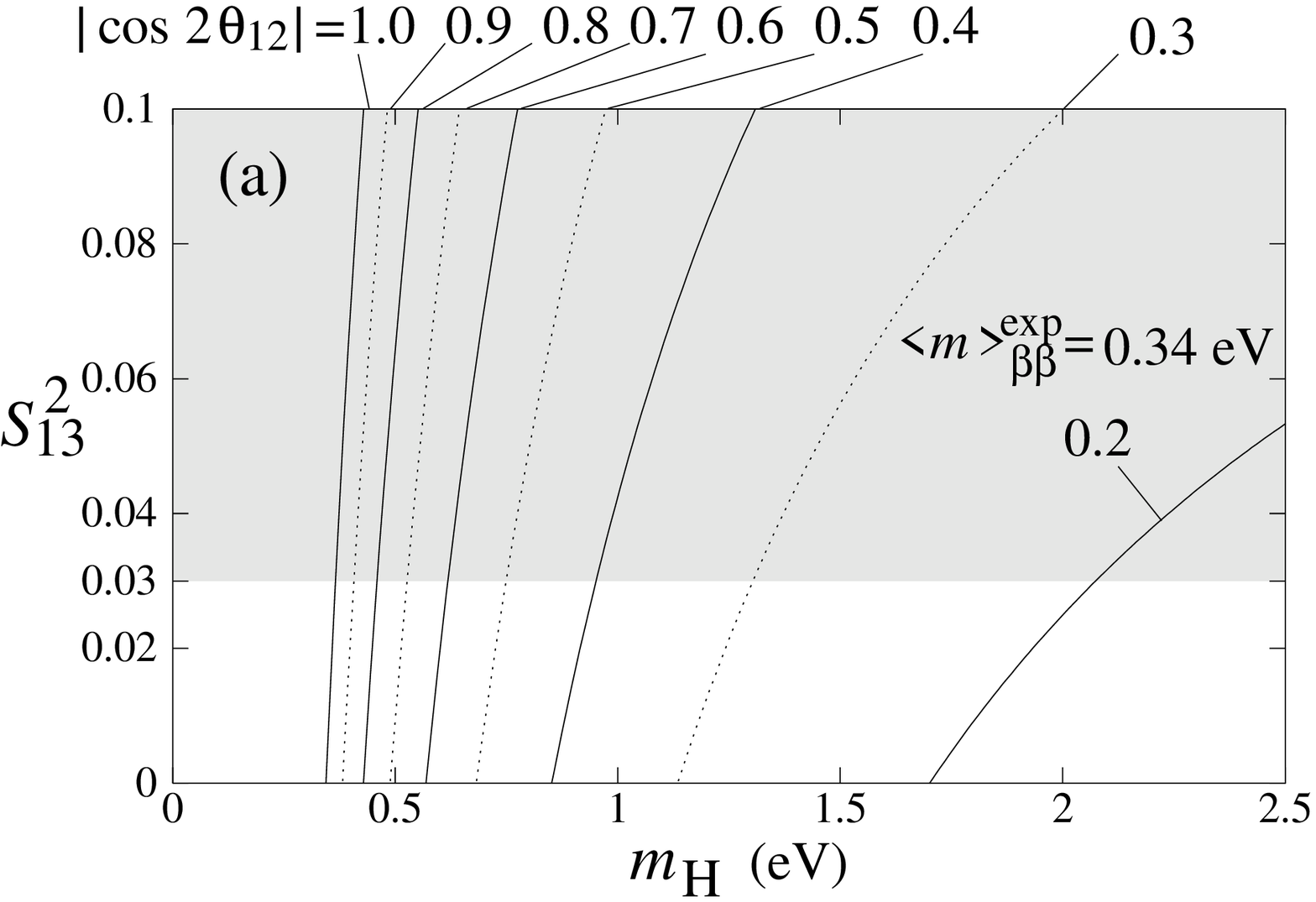,height=7.5cm,width=12cm}
}}
\vglue 1.0cm
\hglue -1.0cm 
\centerline{\protect\hbox{
\psfig{file=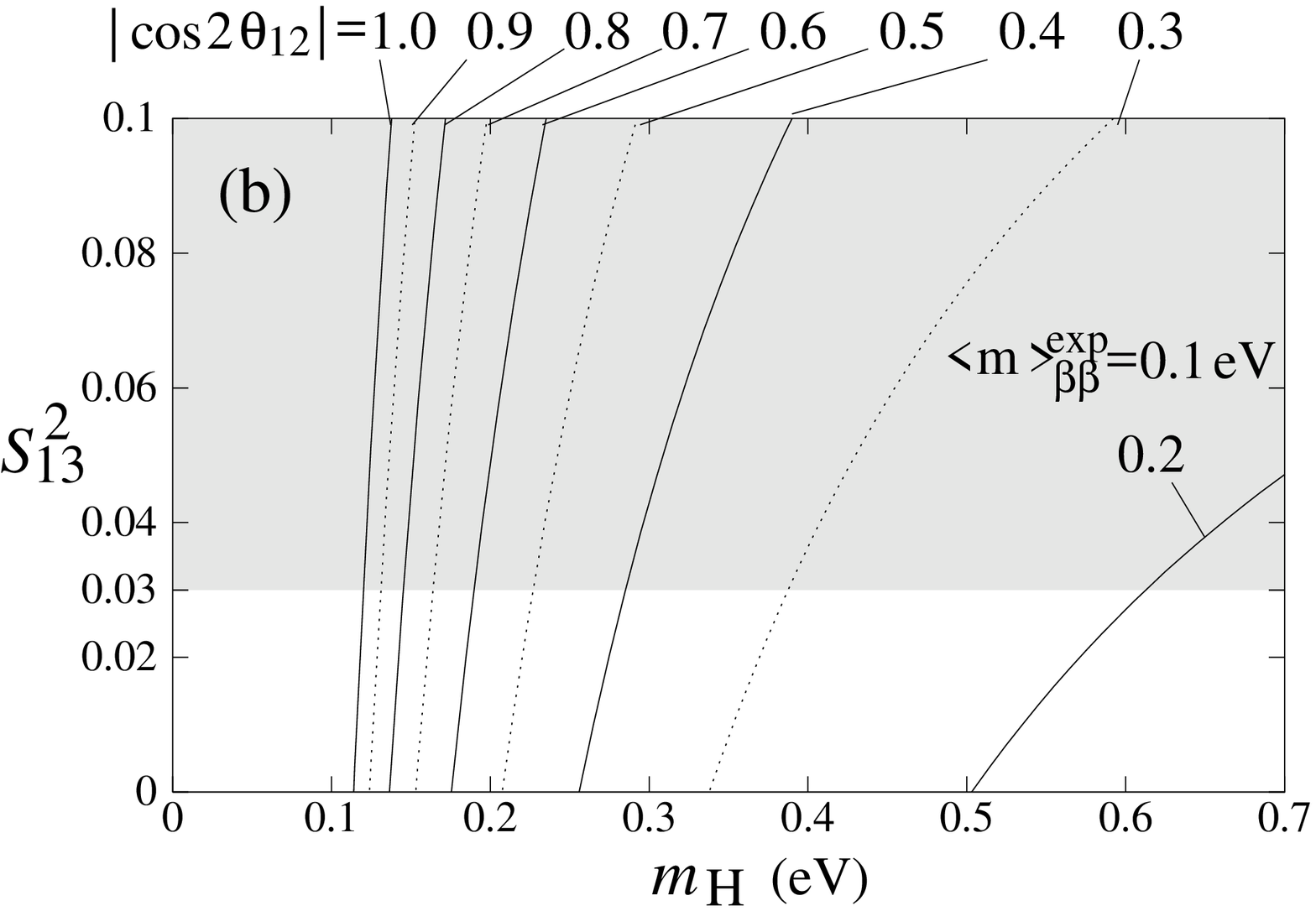,height=7.5cm,width=12cm}
}}
\vglue 1.5cm 
\caption{
The lower bound on $s_{13}^2$ is displayed as a 
function of $m_H$ for various values of $|\cos{2 \theta_{12}}|$ 
as indicated in the figure  
for (a) $\langle m \rangle_{\beta \beta}^{exp} = 0.34$ eV 
and (b) $\langle m \rangle_{\beta \beta}^{exp} = 0.1$ eV. 
The region lower-right to each curve is excluded. 
The shaded area indicates the region excluded by the CHOOZ 
experiment. Only the case of normal mass hierarchy is presented. 
(See the text.)
}
\label{Fig1}
\end{figure}

\newpage

\begin{figure}[ht]
\vglue 2.0cm 
\hglue -1.0cm 
\centerline{\protect\hbox{
\psfig{file=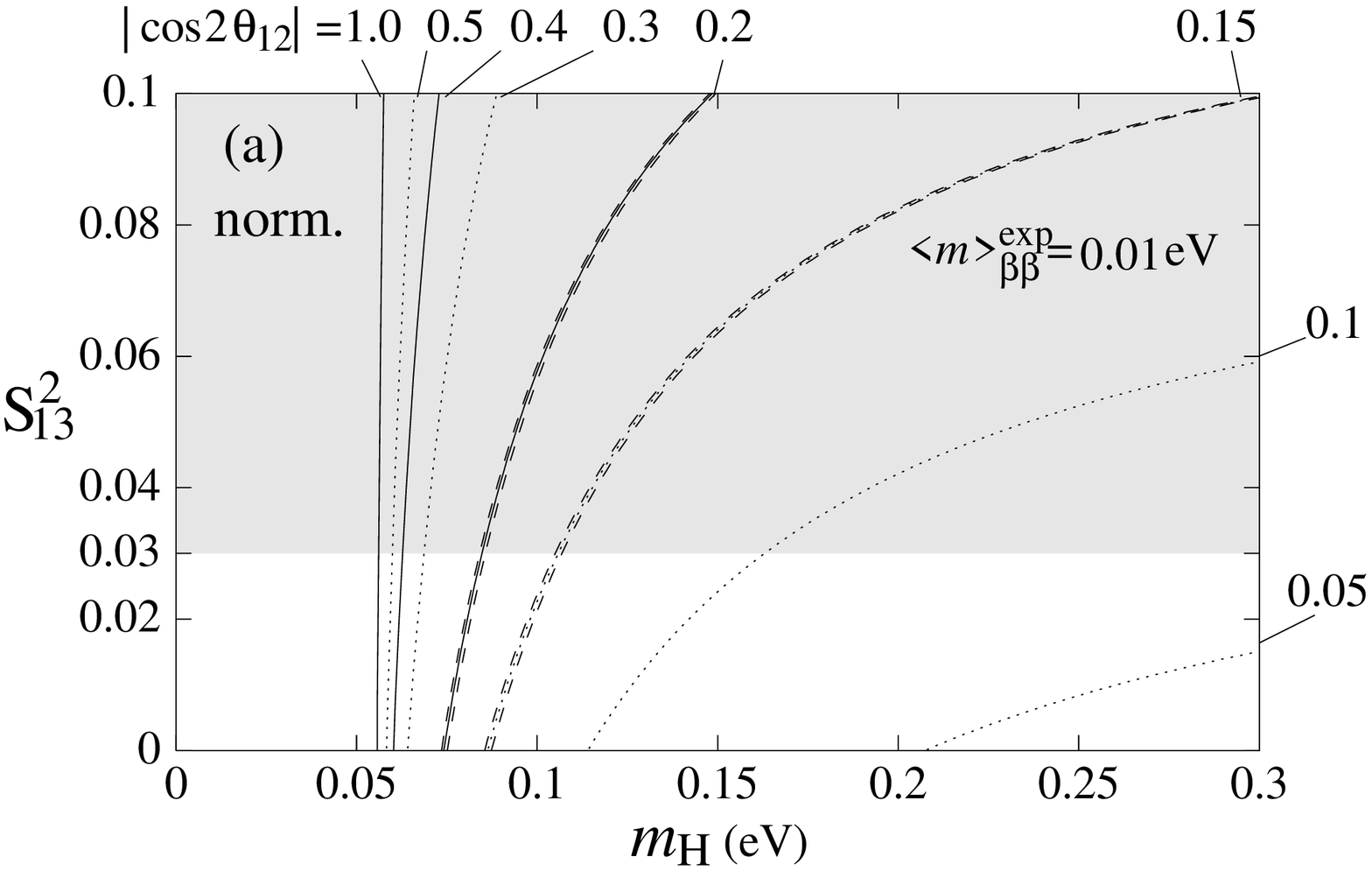,height=7.0cm,width=12.cm}
}}
\vglue 1.0cm
\hglue -1.0cm 
\centerline{\protect\hbox{
\psfig{file=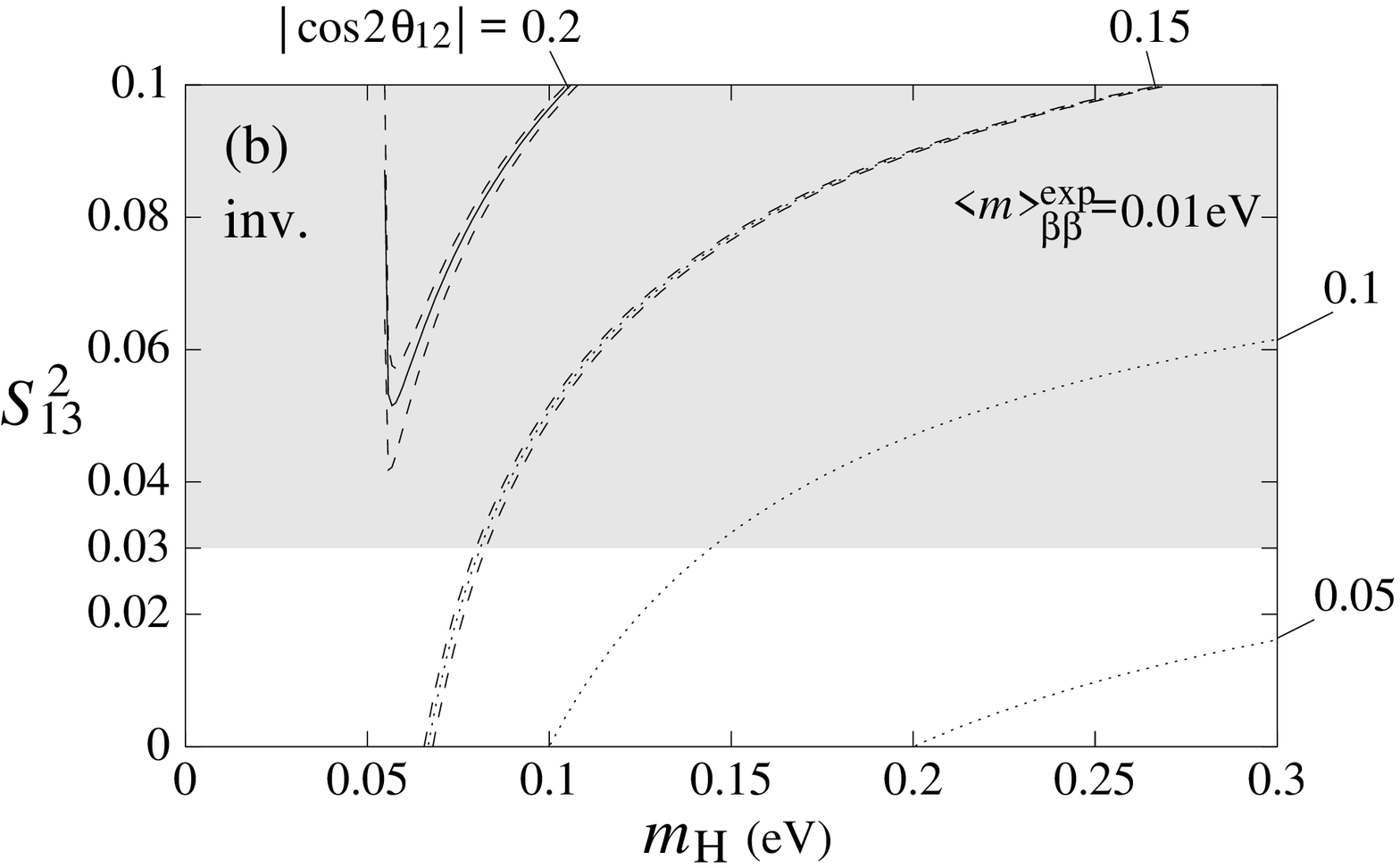,height=7.0cm,width=12.cm}
}} 
\vglue 1.5cm 
\caption{
The same as in Fig. 1 but with 
$\langle m \rangle_{\beta \beta}^{exp} = 0.01$ eV.
Figures 2a and 2b are for cases of the normal and the 
inverted mass hierarchies, respectively.  
The upper and lower dashed lines around curves of 
$|\cos{2 \theta_{12}}| = 0.2$ and 0.15 in both figures are for 
$\cos{2 \theta_{12}} < 0$ and $\cos{2 \theta_{12}} > 0$ 
cases, respectively, with 
$\Delta m_{\odot}^2 = 4.8 \times 10^{-5}$ eV.}
\label{Fig2}
\end{figure}

\newpage

\begin{figure}[ht]
\vglue 2.0cm 
\hglue -1.0cm 
\centerline{\protect\hbox{
\psfig{file=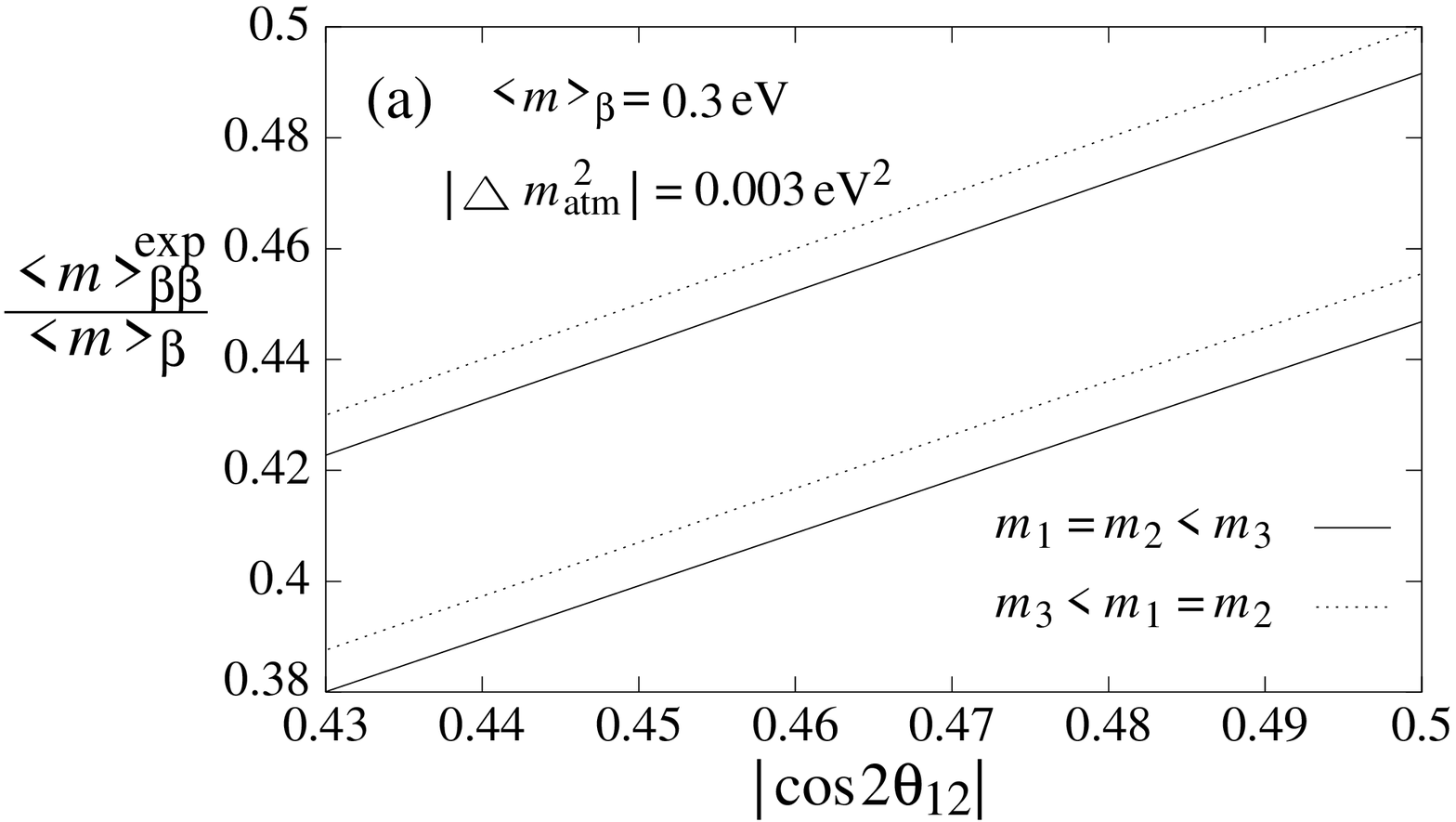,height=7cm,width=12.cm}
}}
\vglue 1.0cm
\hglue -1.0cm 
\centerline{\protect\hbox{
\psfig{file=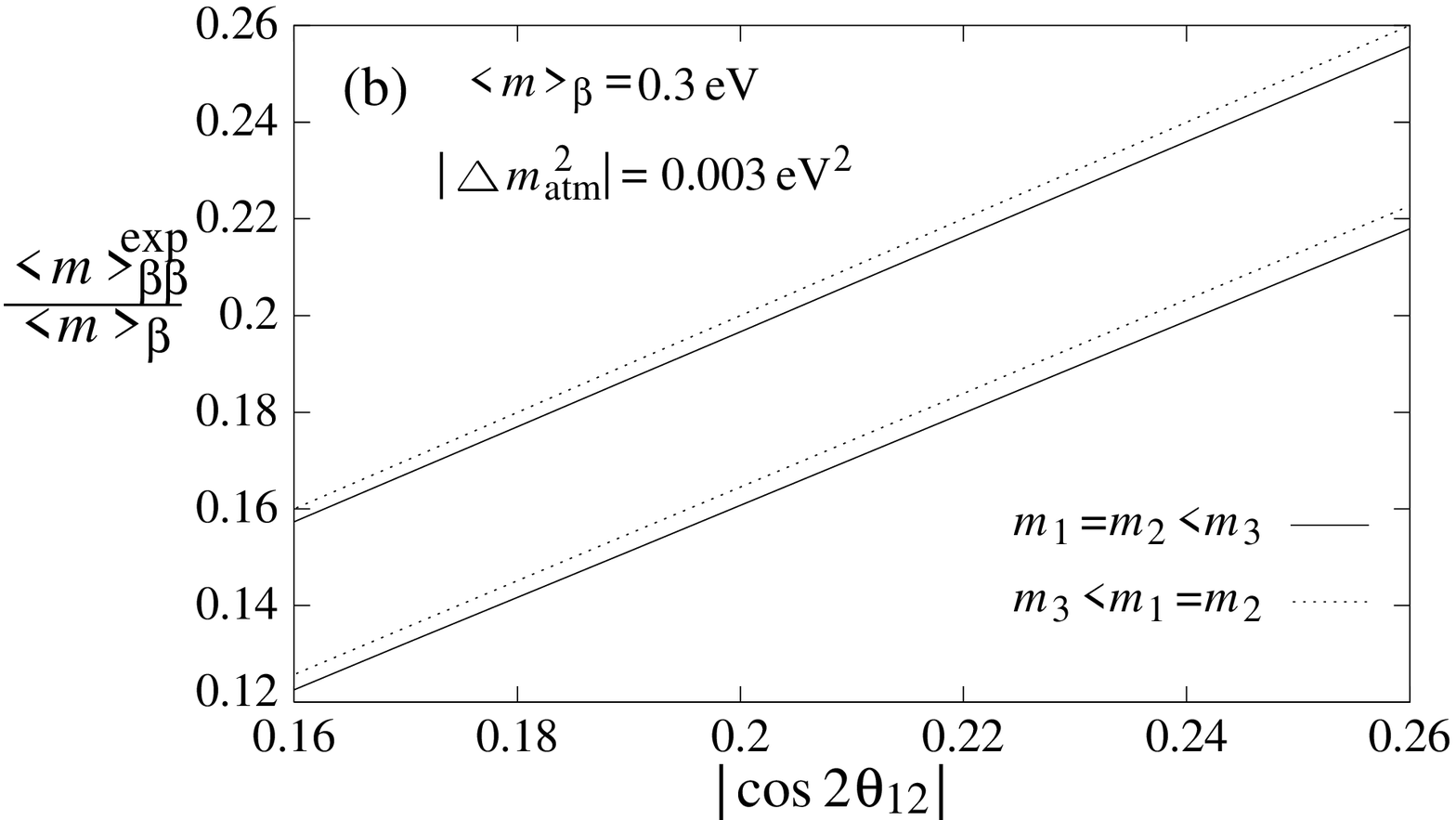,height=7cm,width=12.cm}
}}
\vglue 1.5cm 
\caption{
The region of 
$R_{\beta \beta} \equiv 
\langle m \rangle_{\beta \beta}^{exp}/\langle m \rangle_{\beta}$ 
in which the lower bound exists is exhibited as a region between 
two lines as a function of $|\cos{2 \theta_{12}}|$. 
The solid and dotted lines correspond to the normal and the 
inverted mass hierarchies, respectively. 
Figures 3a and 3b cover roughly the parameter regions of the LMA 
and the LOW solar neutrino solutions, respectively.
}
\label{Fig3}
\end{figure}


\begin{thebibliography}{99}

\bibitem {SKatm}
Kamiokande Collaboration, Y. Fukuda {\it et al.},
Phys. Lett. {\bf B335} (1994) 237;\\
Super-Kamiokande Collaboration, Y. Fukuda {\it et al.},
Phys. Rev. Lett. {\bf 81} (1998) 1562; 
{\it ibid.} {\bf 85} (2000) 3999; \\
T. Kajita, in {\it Neutrino Physics and Astrophysics}, 
Proceedings of the XVIIIth International Conference on Neutrino 
Physics and Astrophysics (Neutrino '98), June 4-9, 1998, Takayama, 
Japan, edited by Y. Suzuki and Y. Totsuka, 
(Elsevier Science B.V., Amsterdam, 1999) page 123.


\bibitem {solar}
Homestake Collaboration, K. Lande {\it et al.},
Astrophys. J.\ {\bf 496} (1998) 505;\\ 
%
SAGE Collaboration, J.\ N.\ Abdurashitov {\it et al.},
Phys.\ Rev.\ C {\bf 60} (1999) 055801; \\
%
GALLEX Collaboration, W.\ Hampel {\it et al.}, Phys.\
Lett.\  B {\bf447} (1999) 127; \\
%
Super-Kamiokande Collaboration,  Y.\ Fukuda {\it et al.}, 
Phys. Rev. Lett. {\bf 86} (2001) 5651; 
{\it ibid.}  {\bf 86} (2001) 5656;\\
%
SNO Collaboration, Q. R. Ahmed {\it et al.}, 
Phys. Rev. Lett. {\bf 87} (2001) 071301.

\bibitem {K2K}
K2K Collaboration, S.~H.~Ahn {\it et al.},
Phys.\ Lett.\ B {\bf 511} (2001) 178;\\
See also http://neutrino.kek.jp/news/2001.07.10.News/index-e.html.

\bibitem {bari}
G. L. Fogli, E. Lisi, and D. Montanino, and A Palazo, 
Phys. Rev. {\bf D64} (2001) 093007.

\bibitem {goswami}
A. Bandyopadhyay, S. Choubey, S. Goswami, and K. Kar, 
Phys. Lett. {\bf B519} (2001) 83.

\bibitem {jnb-concha1}
J. N. Bahcall, M. C. Gonzalez-Garcia, and C. Pe\~na-Garay, 
JHEP {\bf 0108} (2001) 014. 

\bibitem {KS01}
P. I. Krastev and A. Yu. Smirnov, hep-ph/0108177.

\bibitem {jnb-concha2}
J. N. Bahcall, M. C. Gonzalez-Garcia, and C. Pe\~na-Garay, 
hep-ph/0111150. 


\bibitem {CHOOZ}
CHOOZ Collaboration, M.~Apollonio {\it et al.},
Phys.\ Lett.\ B {\bf 420} (1998) 397;
{\it ibid.} B {\bf 466} (1999) 415;\\
The Palo Verde Collaboration,
F.~Boehm {\it et al.},
Phys.\ Rev.\ D {\bf 62} (2000) 072002.


\bibitem {MNS}
Z.~Maki, M.~Nakagawa and S.~Sakata,
Prog.\ Theor.\ Phys.\  {\bf 28} (1962) 870.

\bibitem {NOW2000mina}
H.~Minakata,
Nucl.\ Phys.\ Proc.\ Suppl.\  {\bf 100} (2001) 237.

\bibitem {MINOS}
The MINOS Collaboration, P. Adamson {\it et al.}, 
NuMI-L-337, October 1998.

\bibitem {JHF}
Y. Itow {\it et al.}, 
{\it The JHF-Kamioka neutrino project},
hep-ex/0106019.

\bibitem {OPERA}
OPERA Collaboration, M. Guler {\it et al.}, 
CERN-SPSC-2000-028, CERN-SPSC-P-318, LNGS-P25-00, Jul 2000. 

\bibitem {krasnoyarsk}
Y. Kozlov, L. Mikaelyan, and V. Sinev, hep-ph/0109277.

\bibitem{MNjhep01}
H. Minakata and H. Nunokawa, JHEP {\bf 0110} (2001) 001; hep-ph/0108085.

\bibitem {SNnu}
See, e.g., 
A. S. Dighe and A. Yu. Smirnov, Phys. Rev. {\bf D62} (2000) 033007; \\
H.~Minakata and H.~Nunokawa, Phys.\ Lett.\ B {\bf 504} (2001) 301; \\
K. Takahashi, M. Watanabe, K. Sato, and T. Totani, 
Phys. Rev. {\bf D64} (2001) 093004.

\bibitem {SV}
J. Schechter and J. W. F. Valle, Phys. Rev. {\bf D22} (1980) 2227;\\
S. M. Bilenky, J. Hosek, and S. T. Petcov, Phys. Lett. {\bf B94} 
(1980) 495;\\
M. Doi, T. Kotani, H. Nishiura, K. Okuda, and E. Takasugi, 
Phys. Lett. {\bf B102} (1981) 323.

\bibitem {FY}
M. Fukugita and T. Yanagida, in {\it Physics and Astrophysics of Neutrinos} 
(Springer-Verlag, Tokyo, 1994)

\bibitem {MY97}
H. Minakata and O. Yasuda, Phys. Rev. {\bf D56} (1997) 1692; 
Nucl. Phys. {\bf B523} (1998) 597; 
O. Yasuda, in {\it Proceedings of 2nd International Conference on 
Physics Beyond the Standard Model}, edited by Klapdor-Kleingrothaus 
and I. Krivosheina, pp 223-235 (IOP Bristol 2000).

\bibitem {dblbeta1}
T. Fukuyama, K. Matsuda, and H. Nishiura, 
Mod. Phys. Lett. {\bf A13} (1998) 2279;
Phys. Rev. {\bf D57} (1998) 5844; 
{\it ibid.} {\bf D62} (2000) 093001; {\bf D63} (2000) 077301; 
{\bf D64} (2001) 013001; 
F. Vissani, JHEP {\bf 9906} (1999) 022;\\
V. Barger and K. Whisnant Phys. Lett. {\bf B456} (1999) 194;\\
J. Ellis and S. Lola, Phys. Lett. {\bf B458} (1999) 310;\\
G. C. Branco, M. N. Rebelo, and J. I. Silva-Marcos, 
Phys. Rev. Lett. {\bf 82} (1999) 683;\\
S. M. Bilenky, C. Giunti, W. Grimus, B. Kayser, and S. T. Petcov, 
Phys. Lett. {\bf B465} (1999) 193;\\
M. Czakon, J. Gluza, and M. Zralek, Phys. Lett. {\bf B465} (1999) 211;\\
H. Georgi and S. L. Glashow, Phys. Rev. {\bf D61} (2000) 097301;\\
R. Adhikari and G. Rajasekaran, Phys. Rev. {\bf D61} (2000) 031301. 


\bibitem {dblbeta2}
H. V. Klapdor-Kleingr\"othaus, H. P\"as and A. Yu. Smirnov, 
Phys. Rev. {\bf D63} (2001) 073005; \\
S. M. Bilenky, S. Pascoli, and S. T. Petcov, 
Phys. Rev. {\bf D64} (2001) 053010;  \\
S. Pascoli, S. T. Petcov, and L. Wolfenstein, hep-ph/0110287.

\bibitem {FLMP}
G.~L.~Fogli, E.~Lisi, D.~Montanino and A.~Palazzo,
Phys.\ Rev.\ D {\bf 62} (2000) 013002.


\bibitem {sub-eV}
International Workshop on Neutrino Masses in the Sub-eV Range, 
Bad Liebenzell, Germany, January 18-21, 2001;
http://www-ik1.fzk.de/tritium/liebenzell/.

\bibitem {FPS01}
Y. Farzan, O. L. G. Peres, and A. Yu. Smirnov, 
Nucl. Phys. {\bf B612} (2001) 59.

\bibitem {klapdor}
H. V. Klapdor-Kleingr\"othaus, 
Nucl.\ Phys.\ Proc.\ Suppl.\  {\bf 100} (2001) 350.

\bibitem {CUORE}
E. Fiorini {\it et al.} Phys. Rep. {\bf 307} (1998) 309; 
A. Bettini, Nucl.\ Phys.\ Proc.\ Suppl.\ {\bf 100} (2001) 332.

\bibitem {GENIUS}
GENIUS Collaboration, 
H. V. Klapdor-Kleingr\"othaus {\it et al.}, hep-ph/9910205.

\bibitem {MOON}
H. Ejiri, J. Engel, R. Hazama, P. Krastev, N. Kudomi, and 
R. G. H. Robertson, Phys. Rev. Lett. {\bf 85} (2000) 2919.

\bibitem {Mainz}
J. Bonn {\it et al.}, 
Nucl.\ Phys.\ Proc.\ Suppl.\  {\bf 91} (2001) 273.

\bibitem {Troitsk}
V. M. Lobashev, {\it et al.}, 
Nucl.\ Phys.\ Proc.\ Suppl.\  {\bf 91} (2001) 280.


\bibitem {KATRIN}
KATRIN Collaboration, A. Osipowicz {\it et al.}, hep-ex/0109033.

\bibitem {FLM01}
G. L. Fogli, E. Lisi, and A. Marrone, hep-ph/0110089.


\bibitem {MP01}
H. Murayama and A. Pierce, Phys. Rev. {\bf D65} (2002) 013012. \\
See also, V. Barger, D. Marfatia, and B. P. Wood, 
Phys. Lett. {\bf B498} (2001) 53; \\
A. de Gouv\^ea and C. Pe\~na-Garay, Phys. Rev. {\bf D64} (2001) 113011.

\bibitem {MS02}
H. Minakata and H. Sugiyama, in preparation. 

\bibitem {fukugita}
M. Fukugita, Talk at Frontiers in Particle Astrophysics and Cosmology; 
EuroConference on Neutrinos in the Universe, Lenggries, Germany, 
September 29 - October 4, 2001.


\end{thebibliography}
\end{document}